\documentclass[pre, twocolumn, floatfix, amsmath, showpacs]{revtex4}

\usepackage{graphicx}
\newcommand{\abs}[1]{\left\vert#1\right\vert}
\newcommand{\avg}[1]{\left\langle#1\right\rangle}

\renewcommand{\Re}{\text{Re}}
\newcommand{\qperp}{\mathbf{q}_\perp}
\newcommand{\qperpabs}{q_\perp}

\newcommand{\ps}[1]{#1\:\mathrm{ps}}
\newcommand{\nm}[1]{#1\:\mathrm{nm}}
\newcommand{\mum}[1]{#1\:\mathrm{\mu m}}
\newcommand{\mm}[1]{#1\:\mathrm{mm}}
\newcommand{\cm}[1]{#1\:\mathrm{cm}}
\newcommand{\mms}[1]{#1\:\mathrm{m^{2}s^{-1}}}

\newcommand{\TiO}{TiO$_2$}
\newcommand{\Ttot}{T_\text{tot}}
\begin{document}

\title{Determination of the diffusion constant using phase-sensitive measurements}
\author{I.M. Vellekoop, P. Lodahl and A. Lagendijk}
\affiliation{Complex Photonic Systems, Faculty of Science and
Technology, University of Twente, P.O.Box 217, 7500 AE Enschede,
The Netherlands}

\date{October 8, 2004}

% ----------------------------------------------------------------
\begin{abstract}
We apply a pulsed-light interferometer to measure both the
intensity and the phase of light that is transmitted through a
strongly scattering disordered material. From a single set of
measurements we obtain the time-resolved intensity, frequency
correlations and statistical phase information simultaneously. We
compare several independent techniques of measuring the diffusion
constant for diffuse propagation of light. By comparing these
independent measurements, we obtain experimental proof of the
consistency of the diffusion model and corroborate phase
statistics theory.
\end{abstract} \maketitle

% ----------------------------------------------------------------
\section{Introduction}
Diffusion is one of the most widely encountered phenomena in
physics. The dissolving of sugar in water, the transfer of heat in
a wire and the transport of carriers in a photodiode are all
examples of diffusion. These processes are all described by the
same diffusion equation. This equation also describes the diffuse
transport of waves in disordered scattering materials. An example
of a diffusing wave is the transport of light through a cloud or a
colloid suspension. Wave diffusion is not limited to light;
acoustic waves, microwaves, quantum particles or even seismic
waves behave completely analogously.

The last couple of decennia wave diffusion has been of strong
interest both from applied as well as fundamental points of view.
In contrast to classical particle diffusion, wave diffusion is
influenced by interference. The recognition that phase plays an
important role in wave diffusion forms the basis for applications
like diffusing wave spectroscopy \cite{Pine-DWS} and optical
coherence tomography \cite{deBoer-tissue-imaging}, which are
invaluable tools in the analysis of colloidal systems and in the
optical imaging of biological tissue. Fundamental interest is
motivated especially by the parallels between light diffusion and
transport of electrons in mesoscopic systems. These parallels have
been demonstrated by the observation of the optical equivalents of
universal conductance fluctuations \cite{Scheffold-UCF} and weak
localization \cite{vAlbada-ebs1, Wolf-ebs}.

Multiply scattering media are characterized by the transport mean
free path $\ell$ (the average distance a wave travels through the
medium before becoming diffuse) and the diffusion constant $D$
(the rate at which diffuse waves spread over the medium). For
electrons $\ell$ can be considerably smaller than the wavelength
$\lambda$ of the electron. When ($\ell\lesssim\lambda/2\pi$),
electrons become localized and the diffusion constant vanishes
\cite{Mott-ioffe-regel, Anderson-em-localization}. This breakdown
of diffusion is called Anderson localization. Anderson
localization of microwaves has been observed in quasi-1D systems
\cite{chabanov-localization-nature}. Observations at optical
wavelengths \cite{Wiersma-localization}, however, remain under
debate \cite{Scheffold-localization-questioned}.

In quasi-1D microwave experiments, localization was shown to have
a distinct effect on the statistical distributions of the
intensity \cite{chabanov-localization-nature} and the phase
\cite{Chabanov-lozalization-delay-time}. Recently it has become
possible to perform dynamic electric field measurements also in
the optical regime, which allow a study of the optical phase
\cite{Kop-time-resolved-setup, Johnson}. These types of
measurements provide a direct measurement of the phase of
diffusing waves and they can give unambiguous proof of the
presence of Anderson localization of light.

Here we report our optical experiments that thoroughly test
wave-diffusion theory by measuring the amplitude and the phase of
light transmitted through a strongly scattering, non-localizing
medium. Using the technique of ultrashort pulse interferometry
\cite{Kop-time-resolved-setup}, we have access to the
time-resolved intensity, the frequency-resolved intensity and the
statistical distribution of the phase delay time. We demonstrate
five different ways of extracting the diffusion constant from this
multitude of experimental data. By comparing the results of these
five different methods, we test the diffusion model thoroughly and
moreover show how to interpret time-resolved and
frequency-resolved measurements consistently.

In Section \ref{sec:theory} of this paper we present a model for
diffusion through a slab. From this model we will derive both the
frequency-dependent and the time-dependent behavior and identify
characteristic parameters that can be extracted from experimental
data. The setup for measuring both amplitude and phase of
transmitted light is described in Section \ref{sec:experiment}. In
Section \ref{sec:results} we present our results and devote
special attention to the comparison of different techniques to
measure the diffusion constant. Our conclusions are given in
Section \ref{sec:conclusion}.

\section{Theory}\label{sec:theory}

\subsection{An exact solution to the diffusion equation}\label{sec:th-frequency}
We consider the diffusion of scalar waves through a slab of random
material. The slab fills the space $0\leq z\leq L$ and is infinite
in the other directions. In this geometry it is convenient to use
Fourier transformed coordinates $\qperp \equiv (q_x, q_y)$ for the
transverse directions. The slab is illuminated from the left
($z<0$) by a pulse at time $t=0$. Since the incident light quickly
loses its directionality due to scattering, it is possible to
model the incoming light by a diffuse source inside the material.
In this paper we assume isotropic scattering. The inclusion of
anisotropic scattering in the source function and the description
of anisotropic diffusion are tremendous, and basically unsolved
complications. At this point we use a source located at a depth
$z_0\approx \ell$ \cite{Akkermans-ebs}. Later we will use a more
sophisticated source. Under these conditions, the ensemble
averaged energy density of diffuse light $I$ is described by the
diffusion equation \cite{vRossum},
\begin{equation}
    \left[\partial_t - D\nabla^2 + D\alpha^2\right] I(\qperp, z; t)
    = \delta(z-z_0)\delta(t)S(\qperp).\label{eq:diffusion}
\end{equation}
In this equation $\alpha\equiv\sqrt{3/(\ell\ell_a)}$ is the
absorption coefficient corresponding to an absorption mean free
path $\ell_a$. The right hand side of Eq.\ \eqref{eq:diffusion} is
the source term, where $S(\qperp)$ describes the transverse
distribution of the source and has the unit of energy. The total
energy in the source pulse is given by $S(\qperp=0)$.

The propagation of light is affected by the boundaries of the
slab. It has been shown \cite{Lagendijk-extrapolation-length,
Zhu-Pine-Weitz} that reflections at the surfaces impose mixed
boundary conditions on the diffusion equation,
\begin{subequations}
\begin{align}
    \partial_{z} I(\qperp, 0; t) &= I(\qperp, 0; t)/z_{e1},\label{eq:BC1}\\
    -\partial_{z} I(\qperp, L; t) &= I(\qperp, L;t)/z_{e2},\label{eq:BC2}
\end{align}
\end{subequations}
where $z_{e1}$ and $z_{e2}$ are so called extrapolation lengths.
In the diffusion model, their values are given by $z_{e1,2} =
2\ell(1+R_{1,2})/3(1-R_{1,2})$. The reflection coefficients $R_1$
and $R_2$, correspond to the left and the right boundaries
respectively. These coefficients can be estimated from Fresnel's
law using the refractive indices of the dielectrics outside of the
slab and the effective index of the random medium
\cite{Zhu-Pine-Weitz}.

We solve the diffusion equation (Eq.\ \eqref{eq:diffusion}) with
mixed boundary conditions analytically in the frequency domain.
This solution can conveniently be used to find the field
correlation function, the total transmission and the average
diffuse traversal time. We use the same approach as in
\cite{Zhu-Pine-Weitz}, with the exception that we extend the model
to allow for different extrapolation lengths at the two boundaries
and use an exponential distribution of the source intensity.

When Eq.\ \eqref{eq:diffusion} is Laplace transformed with respect
to $t$, an expression for the energy density $\tilde{I}$ can be
found directly \cite{Carslaw-Jaeger},
\begin{equation}
    \tilde{I}(\eta, z) = \frac{S(\qperp)}{2D\eta}\left[e^{-\eta\abs{z-z_0}}+A(\eta) e^{\eta z}+B(\eta) e^{\eta (L-z)}\right],\label{eq:intensity-laplace}
\end{equation}
where we have defined
$\eta\equiv\sqrt{i\Omega/D+\qperp^2+\alpha^2}$. The Laplace
transform parameter $\Omega$ describes the frequency of intensity
oscillations and is much smaller than the optical frequency of the
field $\omega$. $A$ and $B$ are found from the boundary conditions
\eqref{eq:BC1} and \eqref{eq:BC2} after tedious algebra
\begin{align}
A(\eta) &= \frac{\gamma^+(z_0) -2[z_{e1}\eta+1]e^{\eta z_0}}{\gamma^-(L)e^{\eta L}},\\
B(\eta) &= \frac{\gamma^+(L-z_0)
-2[z_{e2}\eta+1]e^{\eta(L-z_0)}}{\gamma^-(L)e^{\eta L}},
\end{align}
where we defined function $\gamma^\pm$ as
\begin{equation}
\gamma^\pm (x) \equiv (z_{e1}\eta+1)(z_{e2}\eta+1) e^{\eta x}\pm
(z_{e1}\eta-1)(z_{e2}\eta-1) e^{-\eta x}.
\end{equation}
We now have the exact solution to the diffusion equation with
mixed boundary conditions. In order to find the transmitted
intensity flux, we calculate the forward flux $\tilde{J}_z =
-D\partial_z \tilde{I}$ at the slab surface $z=L$,
\begin{equation}
    \tilde{J}_z(\eta, L) = S(\qperp)\frac{F_{z_0}(\eta) - F_{z_0}(-\eta)}
    {\gamma^-(L)},\label{eq:J-solution}
\end{equation}
where $F_{z_0}$ is given by
\begin{equation}
    F_{z_0}(\eta) \equiv [z_{e1}\eta+1] e^{\eta z_0}.\label{eq:F-sheet}
\end{equation}

Eq.\ \eqref{eq:J-solution} describes the transmission for a source
located at depth $z_0$. A more realistic and more sophisticated
model assumes an exponential distribution of the source light. The
exponential distribution models how light becomes diffuse by being
scattered out of the incoming coherent beam. We adapt Eq.\
\eqref{eq:J-solution} for the exponential source model by
convolving $F_{z_0}$ with a (normalized) exponential source
function,
\begin{align}
    F_\ell(\eta) &= \int_0^L dz_0
    \frac{\exp(-z_0/\ell)}{\ell[1-\exp(-L/\ell)]}F_{z_0}(\eta; z_0)\\
    &=\frac{1-\exp(L\eta-L/\ell)}{1-\exp(-L/\ell)}\frac{1+z_{e1}\eta}{1-\ell\eta}.\nonumber\label{eq:F-exponential}
\end{align}
We obtain the transmitted flux for the exponential source from
Eq.\ \eqref{eq:J-solution} by simply replacing $F_{z_0}$ by
$F_\ell$.

An important quantity in the analysis of random media is the total
transmission. The total transmission is found by integrating the
flux over the whole back surface of the sample (this corresponds
to taking $\qperp=0$) and integrating over time ($\Omega=0$). The
ensemble averaged total transmission coefficient $\Ttot$ is
therefore defined as
\begin{equation}
    \Ttot \equiv \frac{\tilde{J}_z(\qperp=0, \Omega=0)}{S(\qperp=0)} =
    \frac{\tilde{J}_z(\eta=\alpha)}{S(\qperp=0)}.
\end{equation}
Neglecting absorption, the equation is evaluated to reproduce the
well known result \cite{Rivas-total-transmission}
\begin{equation}
    \Ttot =\frac{\ell+z_{e1}}{L+z_{e1}+z_{e2}}+O(\exp(-L/\ell)).\label{eq:TT}
\end{equation}
This relation between $\Ttot$, $\ell$ and $L$ is often used to
determine the mean free path experimentally by varying $L$.

Next, we calculate the electric field correlation function $C_E$
for the transmitted light. This correlation function contains
information about the dynamics of the diffusion process,
\begin{equation}
    C_E(\Omega)\equiv\frac{\avg{E(\omega)E^*(\omega+\Omega)}}{\avg{|E(\omega)|}\avg{|E(\omega+\Omega)|}}=
    \frac{\tilde{J}_z(\sqrt{i\Omega/D+\alpha^2}, L)}{\tilde{J}_z(\alpha, L)},\label{eq:def-field-correlation}
\end{equation}
where $E(\omega)$ is the complex field amplitude of the
transmitted light for a incoming field of optical frequency
$\omega$ and unit amplitude. The brackets $\avg{\:}$ are used to
explicitly denote ensemble averaging over all possible
configurations of the disordered sample. We obtained the right
hand side by assuming ergodicity and applying the Wiener-Khinchin
theorem.

The field-field correlation function in Eq.\
\eqref{eq:def-field-correlation} is the exact result for diffuse
transport through a slab using mixed boundary conditions and an
exponential source distribution. Earlier results (Refs.
\cite{Zhu-Pine-Weitz} and  \cite{Genack-slab}) are reproduced by
using the simpler sheet source representation of Eq.\
\eqref{eq:F-sheet} and choosing $z_{e1}=z_{e2}$ or
$z_{e1}=z_{e2}=0$.

We will now turn to the intensity correlation function. This
function relates two single channel transmission coefficients. The
single channel transmission coefficient $T$ describes transmission
from one input angle to one output angle. Integrating $T$ over all
outgoing angles yields the total transmission coefficient $\Ttot$.
The intensity correlation function is defined as
\begin{equation}
C_I(\Omega)\equiv \frac{\avg{\delta T(\omega)\delta
T(\omega+\Omega)}}{\avg{T(\omega)}\avg{T(\omega+\Omega)}},\label{eq:CI-def}
\end{equation}
where $\delta T(\omega)\equiv T(\omega)-\avg{T(\omega)}$. A well
known approximation for the intensity correlation function is
given by
\begin{equation}
C_I(\Omega)=|C_E(\Omega)|^2. \label{eq:C1}
\end{equation}
Eq.\ \eqref{eq:C1} is referred to as the $C_1$ approximation and
is valid for diffusive transport in multiply scattering media far
away from the localization transition \cite{Berkovits}. Eqs.\
\eqref{eq:def-field-correlation} and \eqref{eq:C1} show that both
$C_E(\Omega)$ and $C_I(\Omega)$ depend on the diffusion constant
only by means of the reduced frequency $\Omega/D$. Fitting the
frequency dependence of $C_I$ is a commonly used method to extract
the diffusion constant from measured correlation functions.

It is instructive to introduce the characteristic traversal time
for diffusive transmission $\tau_t$ \cite{Landauer}, which is
defined as the average time it takes a pulse of light to travel
through the medium,
\begin{equation}
\tau_t \equiv \frac{\int dt J_z(t, L) t}{\int dt J_z(t,
L)}=-i\lim_{\Omega\rightarrow 0}\frac{\partial
C_E(\Omega)}{\partial
\Omega}.\label{eq:def-diffuse-traversal-time}
\end{equation}
The right hand side was obtained by rewriting the definition of
$\tau_t$ in the Laplace domain representation and using Eq.\
\eqref{eq:def-field-correlation}. For zero absorption we find
\begin{align}
    \tau_t = \frac{L_e^2-6\ell^2-3z_{e1}^2-3z_{e2}^2}{6D}+\frac{z_{e1}^3+z_{e2}^3}{3L_eD}+O(e^{-L/\ell}).\label{eq:diffuse-traversal-time-value}
\end{align}
The diffuse traversal time is of fundamental interest since it
relates to the Thouless criterion for localization
\cite{Thouless-thin-wire}. Furthermore, the time scale is of
practical interest since measuring $\tau_t$ provides a method of
determining the diffusion constant. Our result in Eq.\
\eqref{eq:diffuse-traversal-time-value} gives corrections of order
$z_eL/D$ and higher to the value of $\tau_t = L^2/6D$ found by
Landauer et al. \cite{Landauer}. These corrections are especially
relevant when $L/z_e<10$, which is the case for thin samples or
samples with a high extrapolation length due to internal
reflection.

\subsection{Phase statistics}\label{sec:th-phase}
The crucial difference between diffusion of particles and wave
diffusion is interference. For this reason we are interested in
the phase of light that propagates through a scattering medium.
Analysis of phase information is complementary to the analysis of
the intensity and provides an independent method of measuring the
traversal time $\tau_t$ and therefore the diffusion constant. We
consider only single channel phase statistics, which means that we
relate phase and amplitude for one input angle to the phase and
amplitude for a single output angle.

Since the diffusion equation only describes the average intensity,
an extension is needed in order to predict phase statistics. The
statistical properties of the phase were predicted by van Tiggelen
et al. \cite{vTiggelen-delay-time} by assuming Gaussian statistics
of the transmitted field \footnote{Ref.\
\cite{vTiggelen-delay-time} also gives corrections for
non-Gaussian field statistics. We do not consider these in this
paper.}. This Gaussian assumption is valid when a high number of
independent paths contributes to the field at the back surface of
the random material. The central limit theorem predicts that in
this situation the real and imaginary parts of the fields are
described by a normal distribution \cite{Goodman}. Equivalently,
the field amplitude is Rayleigh distributed and the phase $\phi$
has a uniform distribution between $0$ and $2\pi$. Neither the
distribution of the intensity nor the distribution of the phase
contains information about the diffusion process. Much more
interesting is the probability distribution of the group velocity
delay time $\phi'\equiv d\phi/d\omega$. This probability
distribution reflects dynamic properties of the diffusion process
and provides a method of measuring the diffusion constant. The
statistics of the delay time $\phi'$ were calculated in Ref.\
\cite{vTiggelen-delay-time}. For this calculation the Gaussian
field statistics were extended to describe the correlations of two
fields at nearby frequencies. These correlations are given by the
field-field correlation function. The resulting joint Gaussian
distribution was subsequently used to calculate the probability
distribution of the delay time,
\begin{equation}
    P(\tilde{\phi}') =
    \frac{Q}{2\left[(\tilde{\phi}'-1)^2+Q\right]^{3/2}}, \label{eq:P-dphi}
\end{equation}
where $\tilde{\phi}'\equiv \phi'/\avg{\phi'}$ and $Q$ is a
dimensionless parameter. $\avg{\phi'}$ and $Q$ can be calculated
from the first and second terms in the Taylor expansion of the
field-field correlation function: $C_E = 1 + i\tau_t \Omega - b
\Omega^2 + O(\Omega^3)$, which results in $\avg{\phi'}=\tau_t$ and
$Q\equiv 2b/\tau_t^2-1$ \cite{vTiggelen-delay-time}.

In Ref. \cite{vTiggelen-delay-time} the correlation function for a
system with simplified boundary conditions was used to calculate
$\tau_t$ and $Q$. Here it was shown that without absorption $Q$
equals $2/5$ while with absorption $Q$ is reduced. However, by
carefully examining our solution for mixed boundary conditions,
Eq.\ \eqref{eq:def-field-correlation}, we find that $Q$ increases
above $2/5$ when the extrapolation lengths are nonzero.

The intensity-weighted delay time $W$ is a fundamental quantity
since the sum of this quantity over all incoming and outgoing
angles equals $\pi$ times the density of states in the medium
\cite{Iannaccone-DOS, vTiggelen-delay-time}. The weighted delay
time is defined as
\begin{equation}
    W\equiv T\phi'.\label{eq:def-W}
\end{equation}
$T$ and $\phi'$ are statistically dependent variables; for
channels with a low transmission the probability distribution of
$\phi'$ is broader \cite{Genack-W-corr}. Because of the
statistical dependency, the statistics of the weighted delay time
cannot be deduced from the individual probability distributions of
$T$ and $\phi'$ and has to be calculated on its own. The
probability distribution of $W$ was calculated in a similar way as
the distribution of $\phi'$ and is given by
\cite{vTiggelen-delay-time}:
\begin{equation}
    P(\tilde{W})     = \frac{1}{\sqrt{1+Q}}
    \exp\left(\frac{-2|\tilde{W}|}{\text{sgn}(\tilde{W})+\sqrt{1+Q}}\right),\label{eq:P-W}
\end{equation}
where $\text{sgn}$ is the signum function\footnote{We corrected
the original formula in Ref. \cite{vTiggelen-delay-time} that
contains a Heaviside step function in place of the signum
function.} and $\tilde{W}\equiv W/\avg{W}$. The average weighted
delay time was found \cite{vTiggelen-delay-time} to relate to the
diffuse traversal time according to $\avg{W}=\avg{T}\tau_t$.

The correlation function of the weighted delay time $C_W$ is
defined as:
\begin{equation}
C_W\equiv\frac{\avg{W(\omega)W(\omega+\Omega)}}{\avg{W(\omega)}\avg{W(\omega+\Omega)}}.
\end{equation}
This correlation function was calculated in the $C_1$
approximation (Eq.\ \eqref{eq:C1}) using a joint Gaussian
distribution that relates the fields at four frequencies
\cite{vTiggelen-delay-time}:
\begin{equation}
    C_W(\Omega) = \frac{1}{2\tau_t^2}\left[\abs{\frac{\partial
    C_E(\Omega)}{\partial\Omega}}^2-\Re\left(C_E(\Omega)\frac{\partial^2
    C^*_E(\Omega)}{\partial\Omega^2}\right)\right].\label{eq:WW-correlation}
\end{equation}
Microwave experiments showed that deviations from the $C_1$
approximation ($C_2$ and $C_3$ correlations) cause $C_W$ to decay
with frequency much slower than is described by Eq.\
\eqref{eq:WW-correlation} \cite{Genack-W-corr}. Therefore
measuring $C_W$ provides a good way of testing the validity of the
$C_1$ approximation and of looking for signs of localization.

\subsection{Diffusion in the time domain}\label{sec:th-time}
Although we found an exact solution to the diffusion equation in
the frequency domain, the time-domain behavior is not obvious from
Eq.\ \eqref{eq:intensity-laplace}. In this section we analyze
diffusion in the time domain and we present an alternative
technique for finding the diffusion constant. The time-resolved
transmission can, in principle, be calculated by inverse Laplace
transforming Eq.\ \eqref{eq:J-solution} by means of contour
integration \cite{Complex-variables}. Unfortunately Eq.\
\eqref{eq:J-solution} has an infinite number of poles, none of
which can be found analytically when the extrapolation lengths are
non-zero. Using a different approach we will show that the
diffusion constant can be found by analyzing only the long-time
behavior of the transmitted flux.

A complete set of solutions to the diffusion equation (Eq.\
\eqref{eq:diffusion}) is given by:
\begin{equation}
I_{q_z, \theta}(\qperp, z; t) = \sin(q_z
z+\theta)\exp\left(-[q^2+\alpha^2]Dt\right)\Theta(t)\label{eq:I-solutions}
\end{equation}
where $q^2\equiv \qperpabs^2+q_z^2$ and $\Theta(t)$ is the
Heaviside step function. In an infinite medium the longitudinal
spatial frequency $q_z$ and phase $\theta$ can be chosen freely.
In a finite slab, however, there is an infinite, discrete set of
combinations of $q_z$ and $\theta$ for which the boundary
conditions are fulfilled. For the boundary conditions given by
Eqs.\ \eqref{eq:BC1} and \eqref{eq:BC2}, permitted values of $q_z$
and the corresponding $\theta$ can be calculated numerically.
Every solution for $q_z$ corresponds to two poles in Eq.\
\eqref{eq:J-solution} with $\eta=\pm i q_z$.

We will only calculate the long-time behavior of diffusion. In the
long-time limit only the solution with the lowest $q_z$ survives,
since according to Eq.\ \eqref{eq:I-solutions} all other solutions
decay faster. We number this particular solution $q_{z1},
\theta_1$. Now we are able to calculate the diffuse flux for
$t\gg1/q^2_{z1}D$,
\begin{equation}
J_z(\qperp, z; t) =-J_0(\qperp)\cos(q_{z1} z+\theta_1)
\exp\left(-[q_{z1}^2+\alpha^2]Dt\right)\label{eq:J-long-time}
\end{equation}
$J_0$ can be calculated by contour integrating Eq.\
\eqref{eq:J-solution} around the poles at $\eta=\pm i q_{z1}$. In
this article we are interested only in the exponential decay time
of the transmitted flux and therefore will not explicitly specify
$J_0$. For the total flux ($\qperp=0$) we find an exponential
decay with a decay time $\tau_d$,
\begin{subequations}
\begin{align}
\tau_d^{-1} &=
\left[q_{z1}^2+\alpha^2\right]D,\label{eq:decay-time}\\
&\approx\left[\frac{\pi^2}{L_e^2}+\alpha^2\right]D,\label{eq:decay-time-approx}
\end{align}
\end{subequations}
where the approximate solution in Eq.\
\eqref{eq:decay-time-approx} was found by linearly extrapolating
$I(\qperp, z; t)$ at the slab boundaries (this is equivalent to
the method of mirror images used in Ref. \cite{vdMark-thesis,
Genack-slab}) and $L_e \equiv L+z_{e1}+z_{e2}$ is an effective
slab thickness. The approximate solution (Eq.\
\eqref{eq:decay-time-approx}) can be used for thick samples ($L\gg
z_{e1}, z_{e2}$).

It is interesting to notice the differences between the decay time
$\tau_d$ and the diffuse traversal time $\tau_t$. The decay time
$\tau_d$ describes the long-time decay rate of the energy density
of diffuse light in the sample. This decay rate is given by the
slowest term in Eq.\ \eqref{eq:I-solutions} and does not depend on
the distribution of the source intensity. The diffuse traversal
time, on the other hand, has contributions from all terms in Eq.\
\eqref{eq:I-solutions} and is mainly determined by the short-time
transmission. The diffuse traversal time does depend on the
distribution of the source intensity. Concluding, $\tau_d$ and
$\tau_t$ are time scales that correspond to different aspects of
diffusion. Therefore, the consistency of the diffusion model can
be tested experimentally by measuring both $\tau_d$ and $\tau_t$
for a series of samples.

\subsection{Apparent non-exponential decay in a realistic experimental configuration}
In the previous section we found that the total transmitted flux
decays single exponentially in the long-time limit. In an actual
experimental geometry, however, it is not possible to collect all
the transmitted light; only a finite area at the back surface of
the sample can be imaged on the detector. We model the limited
area by means of a Gaussian detection efficiency with a known
waist $w_d$. Furthermore, we assume that the source light,
$S(\qperp)$, has a Gaussian intensity distribution with waist
$w_s$. The total intensity reaching the detector $J_\text{det}$ is
found by integrating over all spatial frequencies $\qperp$,
\begin{equation}
    J_\text{det}(t) = \frac{\pi w_d^2}{2} \int d\qperp J_z(\qperp; L, t) \exp(-\frac18 \qperpabs^2 w_d^2),\label{eq:focus-flux}
\end{equation}
The intensity profile at the sample surface is time dependent
according to Eq.\ \eqref{eq:I-solutions} since modulations with a
high spatial frequency $\qperpabs$ decay faster than those with a
low spatial frequency. Since we detect only the flux from a finite
area, our detection efficiency is time dependent as well. We
define $\tau_f\equiv (w_s^2+w_d^2)/8D$, being the characteristic
time scale for the time-dependent detection efficiency, and find
the total detected flux from Eq.\ \eqref{eq:focus-flux},
\begin{equation}
    J_\text{det}(t) = \frac{w_d^2/(8D)}{t+\tau_f} J_z(\qperp=0; L, t).\label{eq:focus-envelope}
\end{equation}
This equation shows that a finite detection area imposes a
non-exponential envelope on the detected transmission and
increases the detected decay rate. For thicker samples the
additional decay will be more pronounced since the diffuse decay,
as described by $\tau_d$, is slower. As a result, the diffusion
constant found from a linear fit of $\ln J(t)$ is structurally
overestimated. Usually the pre-factor in Eq.\
\eqref{eq:focus-envelope} is omitted, corresponding to the
assumption that the detection system collects light from a large
area ($w_d\gg w_s$). The consequences of omitting this correction
can, however, be significant: in our experimental configuration
the correction results in up to a $25\%$ modification of the
measured diffusion constant.

Naturally, the finite-area correction given by
Eq.\eqref{eq:focus-envelope} equally applies for diffusion in the
frequency domain. Unfortunately, it is inconvenient to apply the
correction in the frequency domain analytically. Therefore we use
a numerical fast Fourier transform to correct the
frequency-resolved transmission, Eq.\ \eqref{eq:J-solution}, and
all derived quantities.

\subsection{Five ways of measuring the diffusion constant}\label{sec:th-diffusion-constant}
In Sections \ref{sec:th-frequency}-\ref{sec:th-time} we presented
methods to calculate the frequency correlations, the phase
statistics and the transient behavior of light diffusing through a
slab of randomly scattering material. Two important time scales
were identified: the diffuse traversal time $\tau_t$ and the
exponential decay time $\tau_d$.

In our experiments we will test the consistency of the diffusion
model quantitatively by extracting the diffusion constant from
experimental data using five different techniques. If the model is
valid, we expect all techniques to yield the same diffusion
constant. This will, however, only be the case when the boundaries
and the source intensity distribution are accounted for correctly.
Therefore a comparison of the diffusion constants, measured using
different methods, provides an excellent way of testing our
diffusion model.

\begin{description}
    \item[Method I] In the first method, the diffusion constant is
    found from the diffuse traversal time $\tau_t$. The diffuse traversal time is obtained from time-resolved transmission using
    the definition in Eq.\ \eqref{eq:def-diffuse-traversal-time}. After applying the finite-area correction, the
    diffusion constant is found by means of Eq.\
    \eqref{eq:diffuse-traversal-time-value}. Since the transmitted intensity decays exponentially
    the value of $\tau_t$ depends mainly on the transmission at short time scales.

    \item[Method II] The second method is to measure the decay time $\tau_d$ by fitting the long-time decay of the transmitted
    flux. Subsequently, Eq.\ \eqref{eq:decay-time} is used to find the diffusion constant.
    Since Method II relies on the time-resolved transmission at long time
    scales, the parts of the data used in Method I and Method II are nearly independent.

    \item[Method III] In the third method, the intensity
    correlation function is extracted from frequency-resolved
    measurements. Fitting Eq.\ \eqref{eq:C1} to the measured
    correlation function yields the diffusion constant.

    \item[Method IV] The fourth method relies on the measured optical phase and
    makes use of the statistics derived for the phase of diffuse
    light. When the field obeys Gaussian statistics $\avg{\phi'}$ equals the diffuse traversal time $\tau_t$. Consequently, Eq.\
    \eqref{eq:diffuse-traversal-time-value} can be used to extract the diffusion constant from the measured
    phase. Since Method IV only uses phase information and Methods III only uses the measured intensity, these two methods are fully independent.

    \item[Method V] In the last method the diffusion constant is extracted from measurements of the weighted delay time. With
    $\avg{W}/\avg{T}=\tau_t$ we find the diffuse traversal time.
    As in Methods I and IV we calculate the diffusion constant using Eq.\ \eqref{eq:diffuse-traversal-time-value}.
    It has been shown \cite{Sebbah-delay-time} that
    Method V is mathematically equivalent to Method I. Therefore, we will only
    use Method V to verify the consistency of our data processing.
\end{description}

All together we now have five different methods of measuring the
diffusion constant. A comparison of the results of these methods
provides a thorough test of the diffusion model and the phase
statistics. Furthermore it enables an unambiguous determination of
the diffusion constant.

\section{Experiment}\label{sec:experiment}
\begin{figure}
  \includegraphics[scale=0.6]{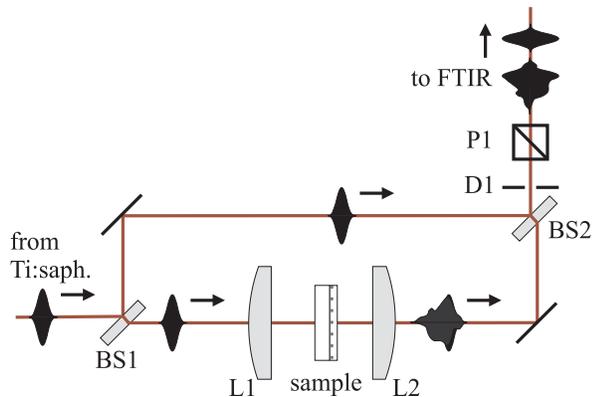}\\
  \caption{Schematic representation of the Mach-Zehnder interferometer used in the
  setup. The first beam splitter (BS1) divides the light between a reference arm and a signal
  arm. The light in the signal arm is focused on the sample by lens
  L1. The transmitted speckle pattern is collimated by lens L2 and recombined with the reference beam at beam splitter
  BS2. Since the reference arm is a few millimeters shorter than the sample arm, the signal pulse does not overlap the reference pulse temporally.
  Finally, aperture $D1$ selects an area that is smaller than the typical speckle size and polarizer P1 blocks light
  with a polarization perpendicular to that of the reference beam in order to increase the signal to noise ratio.
  The beam containing the signal pulse and the reference pulse is
  propagated into a scanning interferometer (FTIR). When a sample is placed in the signal arm,
  only a fraction of the incident light reaches the FTIR. In order to balance the interferometer, we use beam splitters (BS1 and BS2) that reflect approximately 4\%.}\label{fig:setup}
\end{figure}
We have presented a theoretical framework connecting time-resolved
measurements to phase statistics and frequency correlations. In
order to test this framework, we need to measure both the
amplitude and the phase of the multiple scattered light over a
range of optical frequencies simultaneously. We perform these
measurements using the technique of femtosecond pulse
interferometry as described in Ref.
\cite{Kop-time-resolved-setup}. This technique involves two
interferometers. The first interferometer is of the Mach-Zehnder
type and is shown schematically in Figure \ref{fig:setup}. A beam
splitter divides the incoming light between a signal arm and a
reference arm. In the signal arm the light of the laser is
focussed on the sample to a waist diameter of approximately
$\mum{30}$ using a lens with a focal length of $\cm{6}$. In order
to probe different random configurations of scatterers we
illuminate different areas of the sample by translating the sample
perpendicular to the incoming beam. For every sample position the
transmitted light forms a different volume speckle pattern. The
speckle is collimated using a second $\cm{6}$ lens and an area
smaller than a typical speckle spot is selected from the pattern
using an aperture with a diameter of $\mm{0.8}$. At the second
beam splitter the light transmitted through this aperture is
combined with the reference pulse yielding a beam with two
temporally separated pulses.

The double-pulsed signal is directed into a Fourier transform
infrared interferometer (FTIR). The FTIR (Biorad FTS-60A) is a
Michelson interferometer and scans the delay time between two
copies of the signal. A detector directly behind the FTIR obtains
the field autocorrelation function of the pulse pair as a function
of the extra path length in the scanning arm of the
interferometer. Because of the temporal separation of the signal
and reference pulses, it is possible to isolate the
cross-correlate $C(t)$ of the signal pulse with the reference
pulse. In the frequency domain the cross-correlate is given by
\begin{equation}
C(\omega) = |S(\omega)|^2 H_s(\omega)H_r(\omega) E(\omega),
\end{equation}
where $S(\omega)$ is the spectrum of the incoming pulse,
$H_s(\omega)$ and $H_r(\omega)$ are the transfer functions of the
signal and reference arm of the Mach-Zehnder interferometer,
respectively and $E(\omega)$ is the transfer function of the
sample that we wish to extract. In order to find the transfer
function of the sample, the cross-correlate is measured with and
without the sample consecutively. Dividing the two functions
yields the complex transfer function $E(\omega)$ containing both
the phase and the amplitude of the transmitted light. Now the
time-resolved field transmission $E(t)$ can in principle be
obtained by means of an inverse Fourier transform. In practice,
however, the bandwidth of the transfer function is limited by the
bandwidth of the source pulse. Outside this bandwidth the measured
transfer function is dominated by noise, therefore additional
filtering is required before calculating $E(t)$. In our case the
pulses are generated by a Ti:sapphire laser (Tsunami, Spectra
Physics) operating at $\nm{775}$ with a bandwidth of about
$\nm{6}$. We use a Chebyshev filter for filtering in order to have
a minimum effect of side lobes and a maximum time resolution.

In the experiment the signal and the reference beams have to
overlap both at the aperture and at the detector in order to cause
an interference signal. This condition implies that both the
direction and the position of the signal beam are fixed and, as a
result, the detection is limited to light emitted from a small
area of the sample surface. Based on the geometry of the setup we
approximate the detection area by a Gaussian curve with a waist of
$w_d=\mum{10}$.

We perform the measurements on samples consisting of a layer of
rutile \TiO\ particles with a diameter between $\nm{150}$ and
$\nm{290}$ that are deposited on a substrate of fused silica. The
titania grains have a refractive index of approximately $2.8$. The
extrapolation lengths can be calculated from the effective
refractive index $n_\text{eff}$ of the medium
\cite{Zhu-Pine-Weitz}. The effective index can adequately be
estimated from Mie theory \cite{denOuter-Mie}. For our samples we
find $n_\text{eff}=1.34$ and the corresponding extrapolation
lengths are $z_{e1}/\ell=0.69$ for the left boundary and
$z_{e2}/\ell=1.71$ for the right boundary. We measured the total
transmission as a function of sample thickness and found a
transport mean free path of $\ell=\mum{0.97\pm0.10}$ by fitting
the data to Eq.\ \eqref{eq:TT}.

Our samples range in thickness between $\mum{1.5\pm0.3}$ and
$\mum{18.0\pm0.3}$. Since the samples are on a substrate that is
much thicker than the layer of titania, it is necessary to
compensate for the extra delay in the substrate. In order to
accurately determine the extra path length, we direct the light
that is reflected from the substrate into the FTIR without
repositioning the sample. The thickness of the substrate is
deduced from the time delay between the reflections from the front
and the back of the substrate.

With the setup described in this section we are able to measure
the complex transfer function of random media. Below we analyze
these transfer functions in the time domain, the frequency domain
and by looking at the phase statistics.

\section{Results}\label{sec:results}

\subsection{Time domain measurements}\label{sec:res-time}
\begin{figure}
    \includegraphics[scale=0.46]{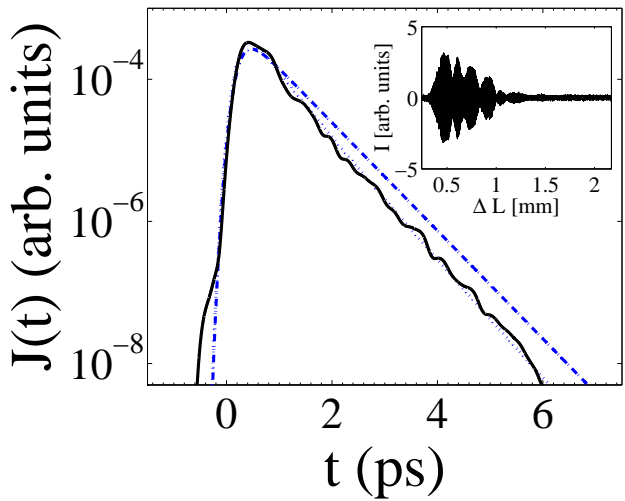}\\
    \caption{Time-resolved intensity transmission for a $\mum{10.1}$ thick sample consisting of
    \TiO\ grains. The observed non-exponential decay (solid line) agrees with the finite-area correction,
    Eq.\ \eqref{eq:focus-envelope}, over four decades.
    Theoretical curves are obtained from Eq.\ \eqref{eq:J-solution} and the shape of the filter.
    The dotted line was corrected for the time-dependent detection efficiency due to
    focussing, using Eq.\ \eqref{eq:focus-envelope} with $w_s=\mum{15}$ and $w_d=\mum{10}$
    as estimated from the experimental configuration and a fitted diffusion constant of
    $D=\mms{27.0}$. The dashed line is the uncorrected curve for the same diffusion
    constant. The inset shows an example of the interference signal at the detector as a function of the delay length in the scanning interferometer.
    The average intensity transmission is obtained from 50 such measurements performed on different areas of the sample.}\label{fig:intensity}
\end{figure}
\begin{figure}
    \includegraphics[scale=0.4]{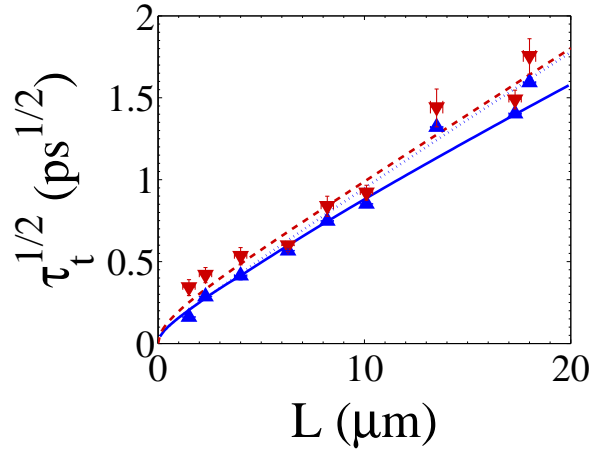}\\
    \caption{Diffuse traversal time $\tau_t$ (triangles pointing upward), and scaled decay time $\tau_d\pi^2/6$ (triangles pointing downward), as a function of sample thickness.
    $\tau_d$ is found by fitting the exponential decay of $J(t)$ (Method II); the error bars indicate the values found from
    fitting the first part of the decay (lowest value) and the last part of the decay (highest
    value). The dashed line is the theoretical value of
    $\tau_d$. The diffuse traversal time is found from the time-resolved intensity transmission
    obtained by numerically evaluating Eq.\ \eqref{eq:def-diffuse-traversal-time} (Method I).
    Without taking into account detection efficiency, theory (dotted
    line) predicts that the two time scales converge for thick samples. When the theory is compensated for the effect of a finite
    detection area (solid line), this convergence is lost in agreement with the experimental data.}\label{fig:slope-time-delay-time}
\end{figure}
First we consider the decay in time of a transmitted pulse. The
inset in Fig.\ \ref{fig:intensity} shows the raw data obtained by
measuring the cross-correlate at a single position of the sample.
For different positions of the sample the transmitted pulse is
distorted differently. We extract $E(t)$ for every measurement as
described in Section \ref{sec:experiment} and average the
corresponding intensities over 50 sample positions to obtain the
normalized transmission $J(t) \equiv \avg{|E(t)|^2}$. Fig.\
\ref{fig:intensity} shows the time-resolved intensity transmission
$J(t)$ for a $\mum{10.1}$ thick sample. We find that in the
long-time limit $J(t)$ has a nearly exponential decay for more
than four decades. The measurements are fitted with the
theoretical curve obtained from Eq.\ \eqref{eq:J-solution}
convolved with the frequency filter that was used in the
processing of the raw data. We find a good fit for a diffusion
constant of $D=\mms{27.0}$ taking into account the effect of the
limited area of detection. For comparison, the theoretical curve
without correction for the detected area is also shown in Fig.\
\ref{fig:intensity}. The corrected curve exhibits a significantly
faster decay, especially for $t<\ps{2}$.

In order to analyze the decay of $J(t)$ more quantitatively, we
extract the diffuse traversal time $\tau_t$ and the decay time,
$\tau_d$ from the measured flux. In Section
\ref{sec:res-diffusion-constant} the diffusion constant will be
calculated from these two times scales using Methods I and II
(Section \ref{sec:th-diffusion-constant}) respectively. The first
time scale $\tau_t$ is obtained from the time-resolved
transmission directly using Eq.\
\eqref{eq:def-diffuse-traversal-time}. The second time scale
$\tau_d$ is extracted from an exponential fit of the intensity
decay. We fit the data between $t>\tau_d$ and the point where the
intensity is dropped below the noise. It was found that the time
constant associated with the first half of this range is always
significantly higher than the time constant for the second half,
as is predicted by Eq.\ \eqref{eq:focus-envelope}. In Fig.\
\ref{fig:slope-time-delay-time} the decay time $\tau_d$ is
compared to the diffuse traversal time $\tau_t$. It was shown in
Section \ref{sec:theory} that the differences between the two time
scales are caused by surface effects and the limited detection
area. We find a good agreement with theory for the decay time as
well as for the traversal time, both using the same fitted average
diffusion constant of $D=\mms{25.5}$.

\subsection{Phase statistics}\label{sec:res-phase}
\begin{figure}
    \includegraphics[scale=0.8]{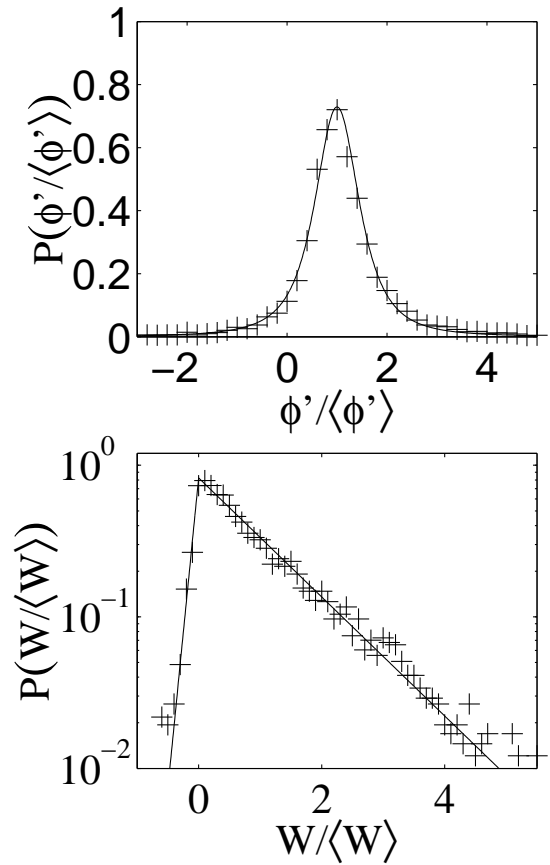}\\
    \caption{Probability distributions for the delay time $\phi'$ (top)
    and the weighted delay time $W$ (bottom) as measured in a $\mum{10.1}$ thick \TiO\ sample.
    The dimensionless parameter $Q$ characterizes the width of
    these distributions. We find $Q=0.47$ from a fit of the theoretical
    curves given by Eqs.\ \eqref{eq:P-dphi} and \eqref{eq:P-W} (solid lines).
    The average values of $\phi'$ and $W$ are used in Methods IV and V
    respectively to find the diffusion constant.}\label{fig:P-dphi-W}
\end{figure}

An independent way of measuring the diffuse traversal time is by
analyzing the phase information. For different positions of the
sample we obtain the phase $\phi$ from the complex transfer
function $E(\omega)$ that was measured using the technique
described in Section \ref{sec:experiment}. For all different
frequencies in the $\nm{6}$ bandwidth of the measurements we
calculate the delay time $\phi'(\omega)\equiv
d\phi(\omega)/d\omega$ and the weighted delay time
$W(\omega)\equiv \phi'(\omega) |E(\omega)|^2$. By binning the
values of $\phi'$ and $W$, we obtain the probability distributions
shown in Fig.\ \ref{fig:P-dphi-W}. The distributions are in good
agreement with the predicted functional forms from theory, Eq.\
\eqref{eq:P-dphi} and Eq.\ \eqref{eq:P-W}. This agreement is a
clear experimental proof that the transmitted light is described
well by a circular complex Gaussian distribution. For a sample
with a thickness of $\mum{10.1}$, the characteristic parameter $Q$
determining the width of the distribution is calculated to be
$Q=0.44$. The experimental data gives $Q=0.47$, corresponding to a
slightly lower maximum of $P(\phi')$. The high value of $Q$
indicates that there is no measurable effect of absorption (which
would decrease $Q$). Moreover, $Q$ is clearly larger than the
value of $2/5$ predicted in Ref.\ \cite{vTiggelen-delay-time}.
This observation shows that even for thick samples ($L\approx
10\ell$) the effect of reflections at the surfaces cannot be
neglected.

\begin{figure}
    \includegraphics[scale=0.4, clip=on]{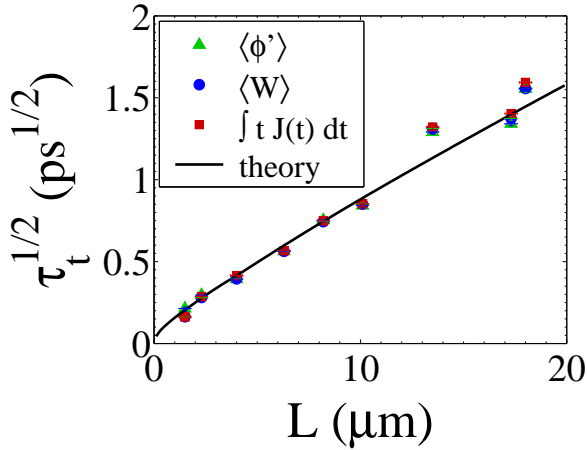}\\
    \caption{Diffuse traversal time $\tau_t$ measured using three different
    techniques. Method I (squares) calculates
    $\tau_t$ from the time-resolved intensity. Method IV (triangles) obtains $\tau_t$ from the measured optical
    phase alone, whereas Method V (circles) uses the intensity-weighted phase information. The excellent agreement indicates that the transmitted field is described by Gaussian distribution.
    The solid line are the theoretical values for a diffusion constant of $D=\mms{25.5}$.}\label{fig:traversal-time-three-times}
\end{figure}

We obtain the diffuse traversal time using $\tau_t=\avg{\phi'}$
(Method IV) and $\tau_t=\avg{W}$ (Method V) and compare these
results to the value found from the time-resolved intensity
measurements (Method I). Fig.\
\ref{fig:traversal-time-three-times} shows $\tau_t$ as obtained by
these three different methods. The values of $\avg{\phi'}$
coincide almost perfectly with $\tau_t$ found from time-resolved
analysis. This agreement again confirms the excellent validity of
the $C_1$ approximation and the theory of phase statistics.

As expected, the results from Methods I and V agree very well.
Although these methods are equivalent in theory, the time-domain
data, on which Method I is based, have been filtered (see Section
\ref{sec:experiment}), whereas Method V uses the unfiltered
frequency-domain measurements directly. Since the differences
between the values obtained by Methods I and V are minute, we
conclude that the determination of $\tau_t$ is insensitive to
frequency domain filtering. In Section
\ref{sec:res-diffusion-constant} we will use $\tau_t$ to extract
the diffusion constant for each sample.

\subsection{Frequency domain measurements}\label{sec:res-frequency}
\begin{figure}
    \includegraphics[scale=0.4]{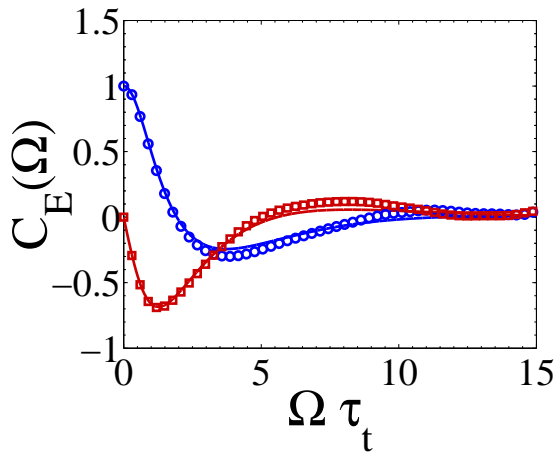}\\
    \caption{Measured field correlation function $C_E$ (real part: circles, imaginary part: squares) for a \TiO\ sample of thickness $\mum{10.1}$ as a function of the frequency difference $\Omega$ between two optical frequencies. The horizontal axis is scaled by $\tau_t^{-1}$ as found
    by measuring $\avg{\phi'}$. Excellent agreement with theory (solid lines) confirms the diffusion model with boundary corrections.}\label{fig:CE-correlation-function}
\end{figure}
\begin{figure}
    \includegraphics[scale=0.4]{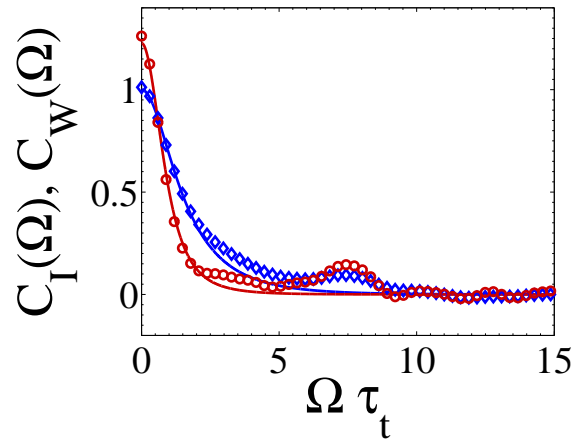}\\
    \caption{Measured correlation functions for the intensity $C_I(\Omega)$ (diamonds), and the weighted delay time $C_W(\Omega)$ (circles), in a $\mum{10.1}$ thick \TiO\ sample.
    The solid lines are the $C_1$ approximations for both correlation functions. Except for some spurious oscillations, agreement with theory is evident.}\label{fig:IW-correlation-function}
\end{figure}
In Sections \ref{sec:res-time} and \ref{sec:res-phase} we
presented measurements of the traversal time $\tau_t$ and $Q$, the
characteristic parameter for phase statistics. These two
parameters are related to the first and second order terms in the
Taylor expansion of the field-field correlation function $C_E$
around $\Omega=0$. In this section we go a step further and
investigate the full frequency correlation functions of the
transmitted light. We investigate the field-field correlation
function, the intensity correlation function and the correlation
function of the weighted delay time consecutively.

We first look at the field-field correlation function. This
function is related to the time-resolved intensity transmission by
a Fourier transform and provides an alternative way of studying
the propagation of diffuse intensity without having to worry about
possible artefacts introduced by filtering. In analyzing the
time-resolved transmission plotted in Fig.\ \ref{fig:intensity} we
only extracted two parameters, $\tau_t$ and $\tau_d$. Whereas the
measured time-resolved transmission curve showed some minor
fluctuations compared to theory, the field-field correlation
function is perfectly smooth up to $\Omega=15\tau_t$ and we find
that the theoretical curve matches the experimental data very
well, as is shown in Fig.\ \ref{fig:CE-correlation-function}.

In order to test the $C_1$ approximation (Eq.\ \eqref{eq:C1})
directly, we examine the intensity correlation function
$C_I(\Omega)$. The presence of long range ($C_2$) and infinite
range ($C_3$) correlations would show up by comparing the
intensity correlation function to the $C_1$ contribution. In our
experiment, however, we find a good agreement to the $C_1$ theory
as is shown in Fig.\ \ref{fig:IW-correlation-function}. At
$\Omega\tau_t\approx 8$ a slight deviation of unknown origin is
found in the correlation function. Surprisingly this deviation was
absent in the field-field correlation function $C_E$. We determine
the diffusion constant by fitting the intensity correlation
function and find a diffusion constant of $\mms{27\pm3}$ for a
$\mum{10.1}$ thick sample.

Finally, we present the correlation function for the weighted
delay time $C_W(\Omega)$ in Fig.\
\ref{fig:IW-correlation-function}. Apart from the same deviations
that were found in the intensity correlation function, the
agreement with Eq.\ \eqref{eq:WW-correlation} is evident. This
observation provides the first experimental confirmation of the
correlation function of the weighted delay time at optical
frequencies.

\subsection{The diffusion constant}\label{sec:res-diffusion-constant}
\begin{figure}
    \includegraphics[scale=0.8]{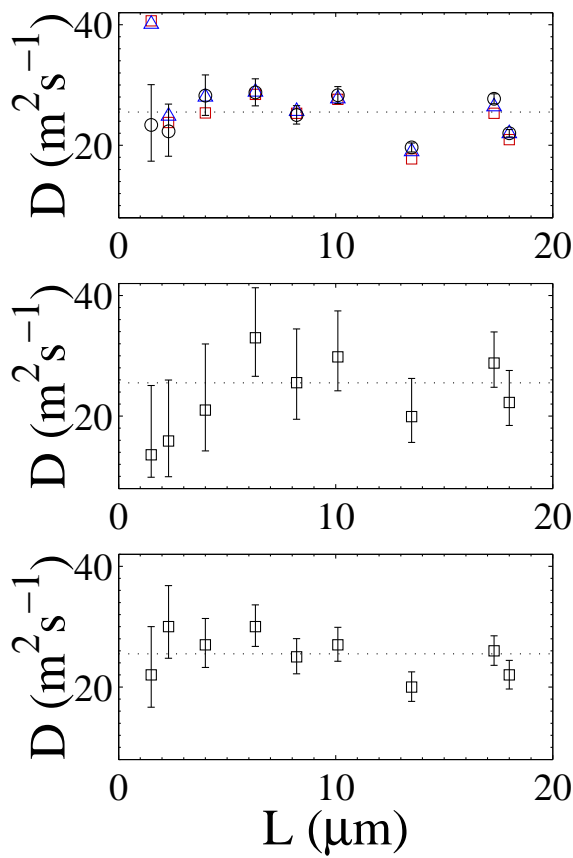}\\
    \caption{Measured diffusion constant for samples of different thicknesses obtained in five different ways.
    The top figure shows the diffusion constant obtained from the diffuse traversal time $\tau_t$.
    We measured the traversal time directly in the time domain (Method I, squares), by extracting the average of the delay time $\avg{\phi'}$ (Method IV, circles),
    and the weighted delay time $\avg{W}$ (Method V, triangles). Error bars are only presented for $\avg{\phi'}$, but the errors in the two other values are comparable. The diffusion constants in the middle figure are calculated
    from the decay time $\tau_d$ which is obtained by fitting the decay of the transmitted flux (Method II). The bottom plot displays the diffusion constants
    obtained from fitting the intensity correlation function (Method III). Except for the two thinnest samples, the consistency of the five different measurement methods is evident.}\label{fig:Diffusion-constants}
\end{figure}
Altogether we have presented five different methods of determining
the diffusion constant experimentally. Methods I and II use the
measured time-resolved intensity transmission to find two time
scales, $\tau_d$ and $\tau_t$, from which the diffusion constant
can be calculated. Subsequently, we showed that $\tau_t$ can be
obtained from phase statistics in two different ways by analyzing
the delay time (Method IV) and the weighted delay time (Method V).
Finally, we measured the diffusion constant by fitting the
intensity correlation function (Method III). The results of these
five methods are summarized in Fig.\ \ref{fig:Diffusion-constants}
for nine samples of different thickness. We find that all
different methods yield the same diffusion constant, within the
experimental accuracy, for a given sample. This observation is an
experimental proof of the consistency of the diffusion model that
was presented in Section \ref{sec:theory}. An average diffusion
constant of $D=\mms{25.5\pm1.0}$ is found; individual samples with
different thickness have slightly different values of the
diffusion constant ranging from $\mms{19}$ to $\mms{28}$. Since
these variations in the diffusion constants are reproduced in for
all methods, we conclude that the scatter is a result of the
varying sample structure and that it is not the result of a
measurement error.

The error bars in Fig.\ \ref{fig:Diffusion-constants} are derived
from the uncertainty in the sample thickness and the uncertainty
in determining the actual value ($\tau_t$, $\tau_d$, or the curve
fit to $C_I$) that was used to calculate the diffusion constant.
For thinner samples we find larger error bars since the
uncertainty in the thickness is relatively large. For the three
methods that are based on the traversal time $\tau_t$ we find that
the uncertainty in $D$ decreases with increasing sample thickness.
The determination of the decay time $\tau_d$ on the other hand,
becomes increasingly more inaccurate because of the
non-exponential decay of the transmitted intensity.

The diffusion constants obtained by fitting the decay of the
transmitted intensity (Method II) appear to be slightly lower for
thinner samples. This behavior is consistent with earlier
observations \cite{Kop-TiO2}. It should be noted, however, that
the decay time in these samples is comparable to the time-domain
resolution. Measurements based on phase information (which are not
limited by the time resolution) show no thickness dependence of
the diffusion constant.

For the thinnest sample we find different diffusion constants
depending on which method we use. The difference is most apparent
when comparing Methods I and V to Method IV. Both the direct
time-domain measurement of $\tau_t$ (Method I) and the average
weighted delay time (Method V) are lower than expected, resulting
in significantly higher values of the measured diffusion constant.
The difference in observed diffusion constants is a clear
indication that the transmitted field does not have a Gaussian
distribution. The discrepancy between the different values of
$\tau_t$ is consistent with an increased transmission at short
times. Therefore this observation suggests an influence of
coherent transmission or single scattering. Especially, it shows
that shorter traversal times are associated with higher
intensities, since the low values of $\tau_t$ are not reproduced
in the unweighted delay time measurements (Method IV).

\section{Conclusion}\label{sec:conclusion}
Pulse interferometric measurements allow a sensitive determination
of the complex transfer function of a random medium. This transfer
function can be used to calculate the time-resolved intensity, the
frequency-resolved intensity and notably the phase of diffuse
light. These three complementary sets of data are obtained in a
single measurement.

In Section \ref{sec:theory} we presented a consistent theoretical
framework to interpret the experimental data. The framework is
built around an exact solution to the diffusion equation with
mixed boundary conditions. Our solution is a generalized version
of the result presented in Ref. \cite{Zhu-Pine-Weitz} and uses a
more accurate description of the source intensity distribution. A
new result is the non-exponential envelope that is imposed on the
detected flux due to a finite detection area. This envelope
contributes to the detected intensity decay and should be taken
into account when extracting the diffusion constant.

By analyzing the diffusion theory both in the time domain and in
the frequency domain we have identified two relevant time scales.
The first is the diffuse traversal time $\tau_t$. The traversal
time is the natural time scale in the analysis of phase statistics
and correlation functions. The second time scale is the decay time
$\tau_d$ associated with the exponential decay of diffuse
intensity. These two times are affected differently by the
boundary conditions and the finite area of detection.

We measured the diffusion constant using five different methods,
which allowed a comparison of time domain measurements, frequency
domain measurements and phase measurements. For our samples
consisting of \TiO\ particles we found that the results of the
five complementary techniques agree almost perfectly for samples
thicker than twice the mean free path. Our observations are a
strong experimental proof of the validity of the diffusion model
and the phase statistics theory.

Finally we measured the correlation function of the weighted delay
time. To our knowledge, this is the first report of such a
measurement at optical wavelengths. The experimental data agree
very well with the measured intensity correlation function and the
predictions from phase statistics theory. These measurements
demonstrate that it is possible to record phase-related
correlation functions at optical wavelengths.

Our analysis clearly shows that care has to be taken in including
proper boundary conditions and correcting for the detection
efficiency, even for samples much thicker than the mean free path.
Provided these effects are taken into account properly, the model
used to describe propagation of light through a random medium is
consistent for all different methods of analysis. Our experiment
demonstrates the versatility and reliability of
pulse-interferometric measurements and validates the use of
phase-sensitive quantities for the identification of long range
correlations and possibly localization of light.

\acknowledgments We thank Boris Bret and Allard Mosk for
stimulating discussions. This work is part of the research program
of the ``Stichting voor Fundamenteel Onderzoek der Materie
(FOM),'' which is financially supported by the ``Nederlandse
Organisatie voor Wetenschappelijk Onderzoek (NWO).''

% ----------------------------------------------------------------
\bibliographystyle{apsrev}
\bibliography{bibliography}

\begin{thebibliography}{32}
\expandafter\ifx\csname natexlab\endcsname\relax\def\natexlab#1{#1}\fi
\expandafter\ifx\csname bibnamefont\endcsname\relax
  \def\bibnamefont#1{#1}\fi
\expandafter\ifx\csname bibfnamefont\endcsname\relax
  \def\bibfnamefont#1{#1}\fi
\expandafter\ifx\csname citenamefont\endcsname\relax
  \def\citenamefont#1{#1}\fi
\expandafter\ifx\csname url\endcsname\relax
  \def\url#1{\texttt{#1}}\fi
\expandafter\ifx\csname urlprefix\endcsname\relax\def\urlprefix{URL }\fi
\providecommand{\bibinfo}[2]{#2}
\providecommand{\eprint}[2][]{\url{#2}}

\bibitem[{\citenamefont{Pine et~al.}(1988)\citenamefont{Pine, Weitz, Chaikin,
  and Herbolzheimer}}]{Pine-DWS}
\bibinfo{author}{\bibfnamefont{D.~J.} \bibnamefont{Pine}},
  \bibinfo{author}{\bibfnamefont{D.~A.} \bibnamefont{Weitz}},
  \bibinfo{author}{\bibfnamefont{P.~M.} \bibnamefont{Chaikin}},
  \bibnamefont{and}
  \bibinfo{author}{\bibfnamefont{E.}~\bibnamefont{Herbolzheimer}},
  \bibinfo{journal}{Phys. Rev. Lett.} \textbf{\bibinfo{volume}{60}},
  \bibinfo{pages}{1134–} (\bibinfo{year}{1988}).

\bibitem[{\citenamefont{de~Boer and Milner}(2002)}]{deBoer-tissue-imaging}
\bibinfo{author}{\bibfnamefont{J.~F.} \bibnamefont{de~Boer}} \bibnamefont{and}
  \bibinfo{author}{\bibfnamefont{T.~E.} \bibnamefont{Milner}},
  \bibinfo{journal}{J. Biomed. Opt.} \textbf{\bibinfo{volume}{7}},
  \bibinfo{pages}{359} (\bibinfo{year}{2002}).

\bibitem[{\citenamefont{Scheffold and Maret}(1998)}]{Scheffold-UCF}
\bibinfo{author}{\bibfnamefont{F.}~\bibnamefont{Scheffold}} \bibnamefont{and}
  \bibinfo{author}{\bibfnamefont{G.}~\bibnamefont{Maret}},
  \bibinfo{journal}{Phys. Rev. Lett.} \textbf{\bibinfo{volume}{81}},
  \bibinfo{pages}{5800} (\bibinfo{year}{1998}).

\bibitem[{\citenamefont{van Albada and Lagendijk}(1985)}]{vAlbada-ebs1}
\bibinfo{author}{\bibfnamefont{M.~P.} \bibnamefont{van Albada}}
  \bibnamefont{and}
  \bibinfo{author}{\bibfnamefont{A.}~\bibnamefont{Lagendijk}},
  \bibinfo{journal}{Phys. Rev. Lett.} \textbf{\bibinfo{volume}{55}},
  \bibinfo{pages}{2692} (\bibinfo{year}{1985}).

\bibitem[{\citenamefont{Wolf and Maret}(1985)}]{Wolf-ebs}
\bibinfo{author}{\bibfnamefont{P.~E.} \bibnamefont{Wolf}} \bibnamefont{and}
  \bibinfo{author}{\bibfnamefont{G.}~\bibnamefont{Maret}},
  \bibinfo{journal}{Phys. Rev. Lett.} \textbf{\bibinfo{volume}{55}},
  \bibinfo{pages}{2696} (\bibinfo{year}{1985}).

\bibitem[{\citenamefont{Mott}(1974)}]{Mott-ioffe-regel}
\bibinfo{author}{\bibfnamefont{N.~F.} \bibnamefont{Mott}},
  \emph{\bibinfo{title}{Metal-Insulator Transitions}}
  (\bibinfo{publisher}{Taylor \& Francis}, \bibinfo{address}{London},
  \bibinfo{year}{1974}).

\bibitem[{\citenamefont{Anderson}(1985)}]{Anderson-em-localization}
\bibinfo{author}{\bibfnamefont{P.~W.} \bibnamefont{Anderson}},
  \bibinfo{journal}{Philos. Mag. B} \textbf{\bibinfo{volume}{52}},
  \bibinfo{pages}{505} (\bibinfo{year}{1985}).

\bibitem[{\citenamefont{Chabanov et~al.}(2000)\citenamefont{Chabanov,
  Stoytchev, and Genack}}]{chabanov-localization-nature}
\bibinfo{author}{\bibfnamefont{A.~A.} \bibnamefont{Chabanov}},
  \bibinfo{author}{\bibfnamefont{M.}~\bibnamefont{Stoytchev}},
  \bibnamefont{and} \bibinfo{author}{\bibfnamefont{A.~Z.}
  \bibnamefont{Genack}}, \bibinfo{journal}{Nature}
  \textbf{\bibinfo{volume}{404}}, \bibinfo{pages}{6780} (\bibinfo{year}{2000}).

\bibitem[{\citenamefont{Wiersma et~al.}(1997)\citenamefont{Wiersma, Bartolini,
  Lagendijk, and Righini}}]{Wiersma-localization}
\bibinfo{author}{\bibfnamefont{D.~S.} \bibnamefont{Wiersma}},
  \bibinfo{author}{\bibfnamefont{P.}~\bibnamefont{Bartolini}},
  \bibinfo{author}{\bibfnamefont{A.}~\bibnamefont{Lagendijk}},
  \bibnamefont{and} \bibinfo{author}{\bibfnamefont{R.}~\bibnamefont{Righini}},
  \bibinfo{journal}{Nature} \textbf{\bibinfo{volume}{390}},
  \bibinfo{pages}{671} (\bibinfo{year}{1997}).

\bibitem[{\citenamefont{Scheffold et~al.}(1999)\citenamefont{Scheffold, Lenke,
  Tweer, and Maret}}]{Scheffold-localization-questioned}
\bibinfo{author}{\bibfnamefont{F.}~\bibnamefont{Scheffold}},
  \bibinfo{author}{\bibfnamefont{R.}~\bibnamefont{Lenke}},
  \bibinfo{author}{\bibfnamefont{R.}~\bibnamefont{Tweer}}, \bibnamefont{and}
  \bibinfo{author}{\bibfnamefont{G.}~\bibnamefont{Maret}},
  \bibinfo{journal}{Nature} \textbf{\bibinfo{volume}{398}},
  \bibinfo{pages}{206} (\bibinfo{year}{1999}).

\bibitem[{\citenamefont{Chabanov and
  Genack}(2001)}]{Chabanov-lozalization-delay-time}
\bibinfo{author}{\bibfnamefont{A.~A.} \bibnamefont{Chabanov}} \bibnamefont{and}
  \bibinfo{author}{\bibfnamefont{A.~Z.} \bibnamefont{Genack}},
  \bibinfo{journal}{Phys. Rev. Lett.} \textbf{\bibinfo{volume}{87}},
  \bibinfo{pages}{233903} (\bibinfo{year}{2001}).

\bibitem[{\citenamefont{Kop and Sprik}(1995)}]{Kop-time-resolved-setup}
\bibinfo{author}{\bibfnamefont{R.~H.~J.} \bibnamefont{Kop}} \bibnamefont{and}
  \bibinfo{author}{\bibfnamefont{R.}~\bibnamefont{Sprik}},
  \bibinfo{journal}{Rev. Sci. Instrum.} \textbf{\bibinfo{volume}{66}},
  \bibinfo{pages}{5459} (\bibinfo{year}{1995}).

\bibitem[{\citenamefont{Johnson et~al.}(2003)\citenamefont{Johnson, Imhof,
  Bret, Rivas, and Lagendijk}}]{Johnson}
\bibinfo{author}{\bibfnamefont{P.~M.} \bibnamefont{Johnson}},
  \bibinfo{author}{\bibfnamefont{A.}~\bibnamefont{Imhof}},
  \bibinfo{author}{\bibfnamefont{B.~P.~J.} \bibnamefont{Bret}},
  \bibinfo{author}{\bibfnamefont{J.~G.} \bibnamefont{Rivas}}, \bibnamefont{and}
  \bibinfo{author}{\bibfnamefont{A.}~\bibnamefont{Lagendijk}},
  \bibinfo{journal}{Phys. Rev. E} \textbf{\bibinfo{volume}{68}},
  \bibinfo{pages}{016604} (\bibinfo{year}{2003}).

\bibitem[{\citenamefont{Akkermans et~al.}(1986)\citenamefont{Akkermans, Wolf,
  and Maynard}}]{Akkermans-ebs}
\bibinfo{author}{\bibfnamefont{E.}~\bibnamefont{Akkermans}},
  \bibinfo{author}{\bibfnamefont{P.~E.} \bibnamefont{Wolf}}, \bibnamefont{and}
  \bibinfo{author}{\bibfnamefont{R.}~\bibnamefont{Maynard}},
  \bibinfo{journal}{Phys. Rev. Lett.} \textbf{\bibinfo{volume}{56}},
  \bibinfo{pages}{1471} (\bibinfo{year}{1986}).

\bibitem[{\citenamefont{van Rossum and Nieuwenhuizen}(1999)}]{vRossum}
\bibinfo{author}{\bibfnamefont{M.~C.~W.} \bibnamefont{van Rossum}}
  \bibnamefont{and} \bibinfo{author}{\bibfnamefont{T.~M.}
  \bibnamefont{Nieuwenhuizen}}, \bibinfo{journal}{Rev. Mod. Phys.}
  \textbf{\bibinfo{volume}{71}}, \bibinfo{pages}{313} (\bibinfo{year}{1999}).

\bibitem[{\citenamefont{Lagendijk et~al.}(1989)\citenamefont{Lagendijk,
  Vreeker, and de~Vries}}]{Lagendijk-extrapolation-length}
\bibinfo{author}{\bibfnamefont{A.}~\bibnamefont{Lagendijk}},
  \bibinfo{author}{\bibfnamefont{R.}~\bibnamefont{Vreeker}}, \bibnamefont{and}
  \bibinfo{author}{\bibfnamefont{P.}~\bibnamefont{de~Vries}},
  \bibinfo{journal}{Phys. Lett. A} \textbf{\bibinfo{volume}{136}},
  \bibinfo{pages}{81} (\bibinfo{year}{1989}).

\bibitem[{\citenamefont{Zhu et~al.}(1991)\citenamefont{Zhu, Pine, and
  Weitz}}]{Zhu-Pine-Weitz}
\bibinfo{author}{\bibfnamefont{J.~X.} \bibnamefont{Zhu}},
  \bibinfo{author}{\bibfnamefont{D.~J.} \bibnamefont{Pine}}, \bibnamefont{and}
  \bibinfo{author}{\bibfnamefont{D.~A.} \bibnamefont{Weitz}},
  \bibinfo{journal}{Phys. Rev. A} \textbf{\bibinfo{volume}{44}},
  \bibinfo{pages}{3948} (\bibinfo{year}{1991}).

\bibitem[{\citenamefont{Carslaw and Jaeger}(1959)}]{Carslaw-Jaeger}
\bibinfo{author}{\bibfnamefont{H.~S.} \bibnamefont{Carslaw}} \bibnamefont{and}
  \bibinfo{author}{\bibfnamefont{J.~C.} \bibnamefont{Jaeger}},
  \emph{\bibinfo{title}{Conduction of heat in solids}}
  (\bibinfo{publisher}{University Press}, \bibinfo{address}{Oxford},
  \bibinfo{year}{1959}), \bibinfo{edition}{2nd} ed.

\bibitem[{\citenamefont{Rivas et~al.}(1999)\citenamefont{Rivas, Sprik,
  Soukoulis, Busch, and Lagendijk}}]{Rivas-total-transmission}
\bibinfo{author}{\bibfnamefont{J.~G.} \bibnamefont{Rivas}},
  \bibinfo{author}{\bibfnamefont{R.}~\bibnamefont{Sprik}},
  \bibinfo{author}{\bibfnamefont{C.~M.} \bibnamefont{Soukoulis}},
  \bibinfo{author}{\bibfnamefont{K.}~\bibnamefont{Busch}}, \bibnamefont{and}
  \bibinfo{author}{\bibfnamefont{A.}~\bibnamefont{Lagendijk}},
  \bibinfo{journal}{Europhys. Lett.} \textbf{\bibinfo{volume}{48}},
  \bibinfo{pages}{22} (\bibinfo{year}{1999}).

\bibitem[{\citenamefont{Genack and Drake}(1990)}]{Genack-slab}
\bibinfo{author}{\bibfnamefont{A.~Z.} \bibnamefont{Genack}} \bibnamefont{and}
  \bibinfo{author}{\bibfnamefont{J.~M.} \bibnamefont{Drake}},
  \bibinfo{journal}{Europhys. Lett.} \textbf{\bibinfo{volume}{11}},
  \bibinfo{pages}{331} (\bibinfo{year}{1990}).

\bibitem[{\citenamefont{Berkovits and Feng}(1994)}]{Berkovits}
\bibinfo{author}{\bibfnamefont{R.}~\bibnamefont{Berkovits}} \bibnamefont{and}
  \bibinfo{author}{\bibfnamefont{S.}~\bibnamefont{Feng}},
  \bibinfo{journal}{Phys. Rep.} \textbf{\bibinfo{volume}{238}},
  \bibinfo{pages}{135} (\bibinfo{year}{1994}).

\bibitem[{\citenamefont{Landauer and B{\"u}ttiker}(1987)}]{Landauer}
\bibinfo{author}{\bibfnamefont{R.}~\bibnamefont{Landauer}} \bibnamefont{and}
  \bibinfo{author}{\bibfnamefont{M.}~\bibnamefont{B{\"u}ttiker}},
  \bibinfo{journal}{Phys. Rev. B} \textbf{\bibinfo{volume}{36}},
  \bibinfo{pages}{6255} (\bibinfo{year}{1987}).

\bibitem[{\citenamefont{Thouless}(1977)}]{Thouless-thin-wire}
\bibinfo{author}{\bibfnamefont{D.~J.} \bibnamefont{Thouless}},
  \bibinfo{journal}{Phys. Rev. Lett.} \textbf{\bibinfo{volume}{39}},
  \bibinfo{pages}{1167} (\bibinfo{year}{1977}).

\bibitem[{\citenamefont{van Tiggelen et~al.}(1999)\citenamefont{van Tiggelen,
  Sebbah, Stoytchev, and Genack}}]{vTiggelen-delay-time}
\bibinfo{author}{\bibfnamefont{B.~A.} \bibnamefont{van Tiggelen}},
  \bibinfo{author}{\bibfnamefont{P.}~\bibnamefont{Sebbah}},
  \bibinfo{author}{\bibfnamefont{M.}~\bibnamefont{Stoytchev}},
  \bibnamefont{and} \bibinfo{author}{\bibfnamefont{A.~Z.}
  \bibnamefont{Genack}}, \bibinfo{journal}{Phys. Rev. E}
  \textbf{\bibinfo{volume}{59}}, \bibinfo{pages}{7166} (\bibinfo{year}{1999}).

\bibitem[{\citenamefont{Goodman}(2000)}]{Goodman}
\bibinfo{author}{\bibfnamefont{J.~W.} \bibnamefont{Goodman}},
  \emph{\bibinfo{title}{Statistical optics}} (\bibinfo{publisher}{Wiley},
  \bibinfo{address}{New York}, \bibinfo{year}{2000}).

\bibitem[{\citenamefont{Iannaccone}(1995)}]{Iannaccone-DOS}
\bibinfo{author}{\bibfnamefont{G.}~\bibnamefont{Iannaccone}},
  \bibinfo{journal}{Phys. Rev. B} \textbf{\bibinfo{volume}{51}},
  \bibinfo{pages}{R4727} (\bibinfo{year}{1995}).

\bibitem[{\citenamefont{Genack et~al.}(1999)\citenamefont{Genack, Sebbah,
  Stoytchev, and van Tiggelen}}]{Genack-W-corr}
\bibinfo{author}{\bibfnamefont{A.~Z.} \bibnamefont{Genack}},
  \bibinfo{author}{\bibfnamefont{P.}~\bibnamefont{Sebbah}},
  \bibinfo{author}{\bibfnamefont{M.}~\bibnamefont{Stoytchev}},
  \bibnamefont{and} \bibinfo{author}{\bibfnamefont{B.~A.} \bibnamefont{van
  Tiggelen}}, \bibinfo{journal}{Phys. Rev. Lett.}
  \textbf{\bibinfo{volume}{82}}, \bibinfo{pages}{715} (\bibinfo{year}{1999}).

\bibitem[{\citenamefont{Brown and Churchill}(1996)}]{Complex-variables}
\bibinfo{author}{\bibfnamefont{J.~W.} \bibnamefont{Brown}} \bibnamefont{and}
  \bibinfo{author}{\bibfnamefont{R.~V.} \bibnamefont{Churchill}},
  \emph{\bibinfo{title}{Complex variables and applications}}
  (\bibinfo{publisher}{McGraw-Hill}, \bibinfo{year}{1996}),
  \bibinfo{edition}{6th} ed.

\bibitem[{\citenamefont{van~der Mark}(1990)}]{vdMark-thesis}
\bibinfo{author}{\bibfnamefont{M.~B.} \bibnamefont{van~der Mark}},
  \emph{\bibinfo{title}{Propagation of light in disordered media: A search for
  Anderson localization}} (\bibinfo{publisher}{University of Amsterdam},
  \bibinfo{address}{Amsterdam}, \bibinfo{year}{1990}).

\bibitem[{\citenamefont{Sebbah et~al.}(1999)\citenamefont{Sebbah, Legrand, and
  Genack}}]{Sebbah-delay-time}
\bibinfo{author}{\bibfnamefont{P.}~\bibnamefont{Sebbah}},
  \bibinfo{author}{\bibfnamefont{O.}~\bibnamefont{Legrand}}, \bibnamefont{and}
  \bibinfo{author}{\bibfnamefont{A.~Z.} \bibnamefont{Genack}},
  \bibinfo{journal}{Phys. Rev. E} \textbf{\bibinfo{volume}{59}},
  \bibinfo{pages}{2406} (\bibinfo{year}{1999}).

\bibitem[{\citenamefont{den Outer and Lagendijk}(1993)}]{denOuter-Mie}
\bibinfo{author}{\bibfnamefont{P.~N.} \bibnamefont{den Outer}}
  \bibnamefont{and}
  \bibinfo{author}{\bibfnamefont{A.}~\bibnamefont{Lagendijk}},
  \bibinfo{journal}{Opt. Comm.} \textbf{\bibinfo{volume}{103}},
  \bibinfo{pages}{169–} (\bibinfo{year}{1993}).

\bibitem[{\citenamefont{Kop et~al.}(1997)\citenamefont{Kop, de~Vries, Sprik,
  and Lagendijk}}]{Kop-TiO2}
\bibinfo{author}{\bibfnamefont{R.~H.~J.} \bibnamefont{Kop}},
  \bibinfo{author}{\bibfnamefont{P.}~\bibnamefont{de~Vries}},
  \bibinfo{author}{\bibfnamefont{R.}~\bibnamefont{Sprik}}, \bibnamefont{and}
  \bibinfo{author}{\bibfnamefont{A.}~\bibnamefont{Lagendijk}},
  \bibinfo{journal}{Phys. Rev. Lett.} \textbf{\bibinfo{volume}{79}},
  \bibinfo{pages}{4369} (\bibinfo{year}{1997}).

\end{thebibliography}
\end{document}